\documentstyle[11pt,paspconf,epsf]{article}
\def\edcomment#1{\iffalse\marginpar{\raggedright\sl#1\/}\else\relax\fi}
\reversemarginpar

\begin{document}
\title{Light Element Evolution in the Galactic Halo}

\author{Andreu Alib\'es}
\affil{Department of Astronomy, University of Barcelona, Mart\'{\i} i Franqu\`es 1, E-08028, Barcelona, Spain. E-mail: aalibes@quasar.am.ub.es}

\author{Ramon Canal}
\affil{Department of Astronomy, University of Barcelona, Mart\'{\i} i Franqu\`es 1, E-08028, Barcelona, Spain. E-mail: ramon@mizar.am.ub.es\\
Institut d'Estudis Espacials de Catalunya/UB, Edif. Nexus-104, Gran Capit\`a 2-4, Barcelona, Spain}

\begin{abstract}

Using a time-dependent Galactic Cosmic Ray flux proportional to the
halo Star Formation Rate and including astration and neutrino-induced
nucleosynthesis, we have studied the evolution of lithium, beryllium and boron
in the halo.  Our results set limits to the production of LiBeB by the neutrino-induced nucleosynthesis in massive stars, in order to reproduce
the observed constancy of Be/Fe and B/Fe ratios, the lithium plateau and the isotopic ratios evolution.
\end{abstract}

\section{Introduction}

Light-element evolution due to spallation reactions between Galactic Cosmic Rays and the Interstellar Medium has been intensively studied in the last thirty years, since the pioneering work of Meneguzzi, Audouze and Reeves (1971). In recent years, lithium, beryllium and boron (LiBeB) abundances have been measured in low-metallicity stars and new constraints have been set on the evolutionary models. The Spite plateau ($T_{eff}>$ 5700 K and $[Fe/H]<$-1.0) for lithium and the [Be] and [B] vs [Fe/H] linear relationship are the main characteristics of LiBeB abundances in halo stars. Our main objective has been to reproduce those observational results and also to get isotopic ratios that agree whith the few data avalaible. In our evolution code, a simple astration model has been used and the possibility of neutrino-induced nucleosynthesis has been considered.

\section{Model}
Using a burst of 1 Gyr as halo star formation rate, we have calculated the age-metallicity relation, including just yields from gravitational supernovae (type Ib and II) (fig. 1). We have not considered yields from type Ia supernovae due to the longer lifetime of their progenitors.

\begin{figure}[h]
\plotfiddle{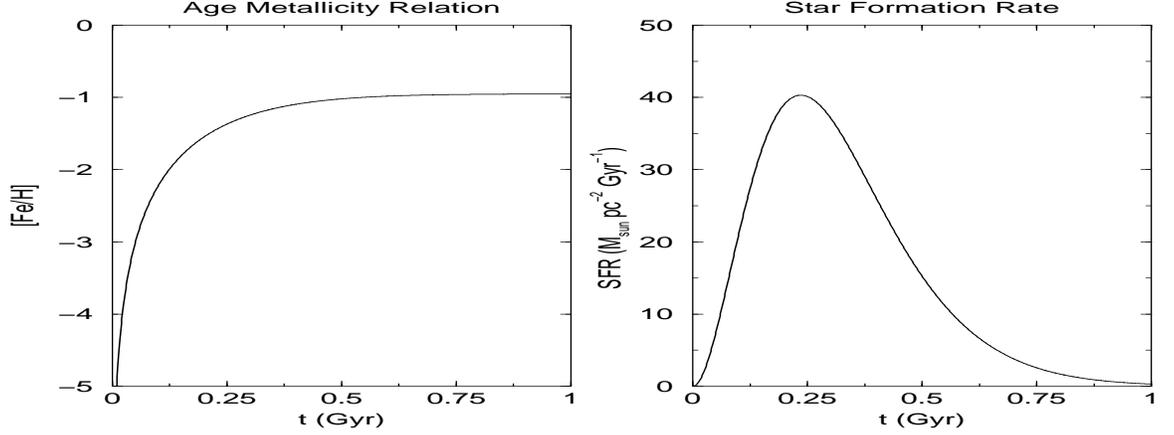}{4cm}{0}{80}{40}{-250}{-10}
\caption{\em Halo star formation rate and halo age-metallicity relation}
\end{figure}
We have used equation (1) for the evolution of the five stable isotopes ($^6$Li, $^7$Li, $^9$Be, $^{10}$B and $^{11}$B)

$$
\frac{d(X_{L}(t)\sigma_g (t))}{dt}=A\ Q_{L}(t)-
X_{L}(t)\int \ IMF(m)\ SFR(t)\ dm+
$$
$$
\int \ f_L (m)\ X_L (t-\tau_m)\ IMF(m)\ SFR(t-\tau_m)\ dm+
$$
\begin{eqnarray}
\int _{12}^{40} \frac{m_{\nu L}(m,[Fe/H])}{m}\ SFR(t-\tau_m)\ IMF(m)\ dm
\end{eqnarray}

\begin{itemize}
\item The first term on the right side represents the contribution of spallation reactions of Galactic Cosmic Rays with the Interstellar Medium. We have considered a Galactic Cosmic Ray flux proportional to the star formation rate, which seems reasonable if we take type II supernovae as the mechanism that accelerates Cosmic Rays. This rate has been calculated with the Ramaty et al. (1997)'s code, using:

\begin{enumerate}

\item the new [O/Fe] vs [Fe/H] data from Israelian et al. (1998)

\item a type II SNe ejecta as composition of  GCR (table 1).

\begin{table}[b]
\caption{\em SNII ejecta composition} \label{tbl-1}
\begin{center}\scriptsize
\begin{tabular}{lcc}
\tableline
& & \\
$\ He/H = 0.20$ & $ \rightarrow $ & $[He/H] = 0.521 $\\
& & \\
$\ C/H = 2.1\cdot 10^{-3}$ & $ \rightarrow $ & $ [C/H] = 0.709 $\\
& & \\
$\ N/H = 5.8\cdot 10^{-4}$ & $ \rightarrow $ & $ [N/H] = 0.825 $\\
& & \\
$\ O/H = 1.2\cdot 10^{-2}$ & $ \rightarrow $ & $ [O/H] = 1.23 $\\
& & \\
\tableline
\tableline
\end{tabular}
\end{center}
\end{table}

\item a shock accelerated spectrum (eq. 2) with $E_0$ = 100 MeV/n
\begin{equation}
q(E)\ \propto \ \frac{p^{-2.2}}{\beta}e^{-E/E_0}
\end{equation}
\item $\Lambda$ = 10 g/cm$^2$, escape length

\end{enumerate}

Other spectra, $E_0$ and escape length have been tried, but those 
finally choosed give the best agreement with the data.

\item The second term represents the amount of each light element that goes into new-born stars. 

\item The third term represents the amount of each light element that each star returns to the interstellar medium after its death. We have used a shell-model star with light elements homogeneously distributed and the reaction rates of Caughlan and Fowler (1988). We have let the star evolve during all its lifetime and then we have obtained the mass fraction of the ejected material where each light nuclide has not been totally destroyed. We represent this function by $f_L(m)$ (fig. 2).

\begin{figure}[h]
\plotfiddle{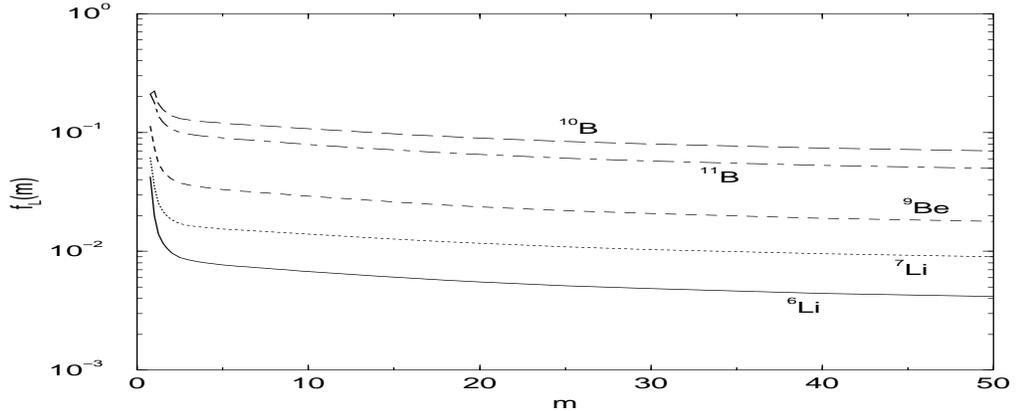}{5cm}{0}{80}{40}{-250}{-30}
\caption{\em Mass fraction of the ejected material where LiBeB have not been totally destroyed}
\end{figure}

\item Finally, the fourth term is the contribution of neutrino-induced nucleosynthesis to the LiBeB evolution.

\end{itemize}

\section{Results}

\subsection{GCR nucleosynthesis alone}
As a first step in our calculations, we didn't include the yields from the neutrino-induced nucleosynthesis in massive stars. The LiBeB evolution that is obtained agrees with the observed evolution: the lithium plateau is well fitted (fig. 3)  and the slope of $1$ in the beryllium and boron versus metallicity relationships is reproduced (fig. 4). 

Concerning light-element ratios, $^6$Li/$^7$Li and Li/B evolution are reproduced and B/Be evolution is close to the average value of 15. However, the evolution of the boron isotopic ratio needs some other production site for $^{11}$B, in order to get at [Fe/H]=-1 a value close to the solar $^{11}$B/$^{10}$B = 4.05 $\pm$ 0.2 (fig. 5).

\begin{figure}[t]
\plotfiddle{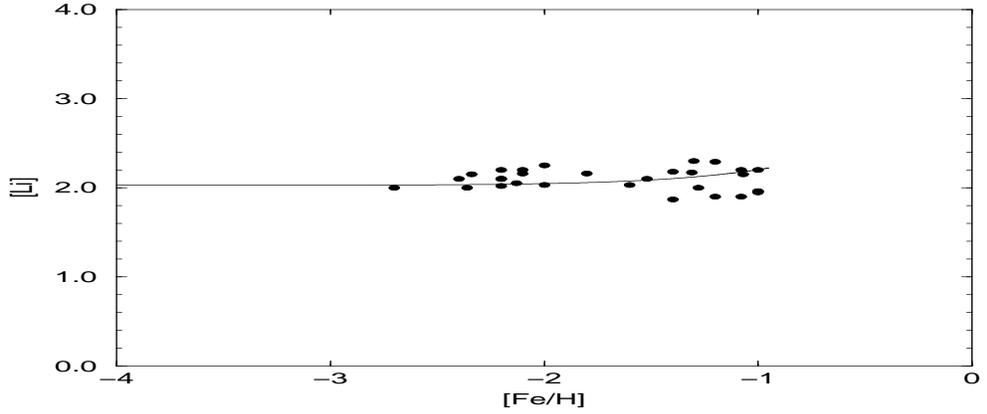}{4cm}{0}{80}{40}{-250}{-30}
\caption{\em Lithium evolution without $\nu$-process. Lithium plateau data: Molaro et al. (1997)}
\end{figure}

\begin{figure}[h]
\plotfiddle{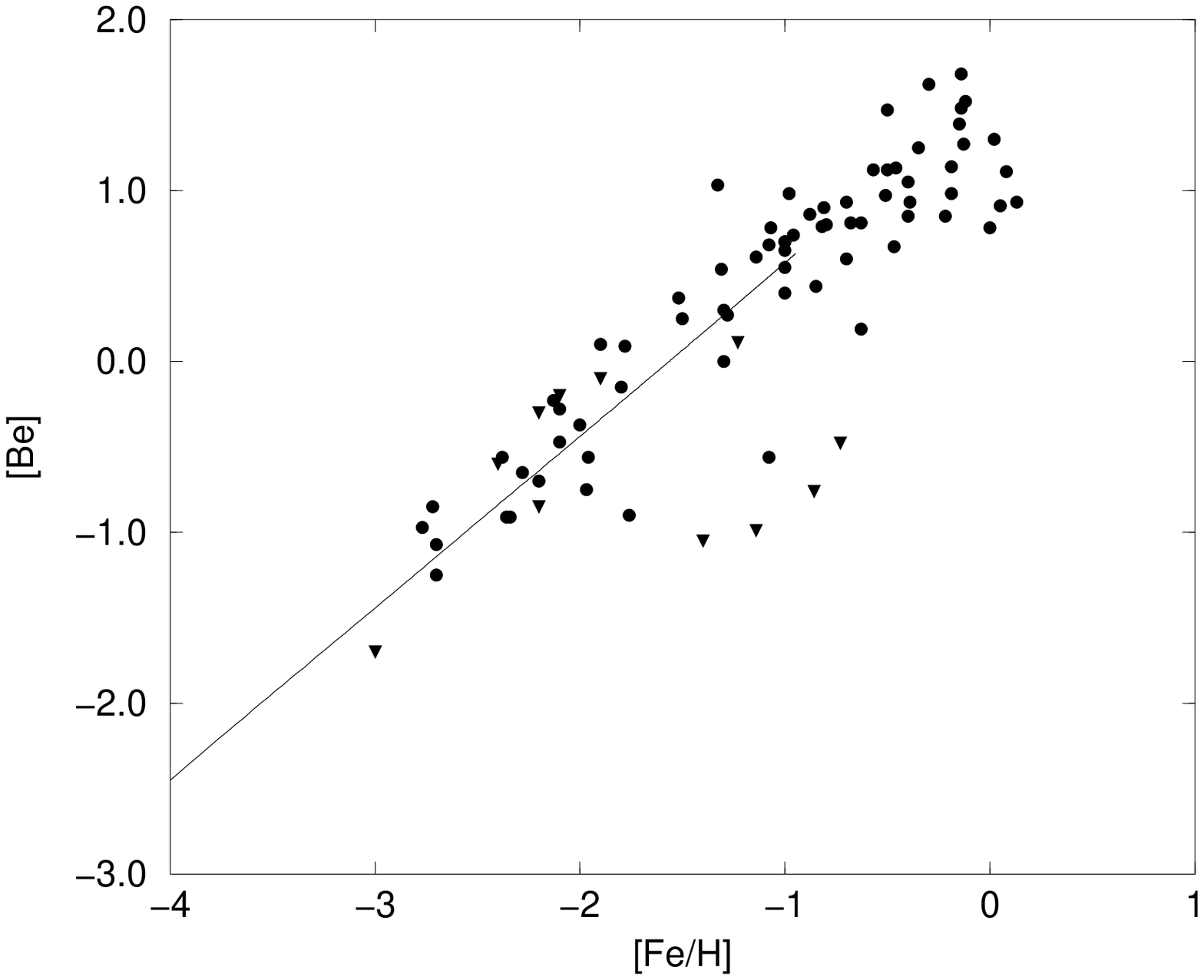}{4cm}{0}{40}{40}{-220}{-50}
\plotfiddle{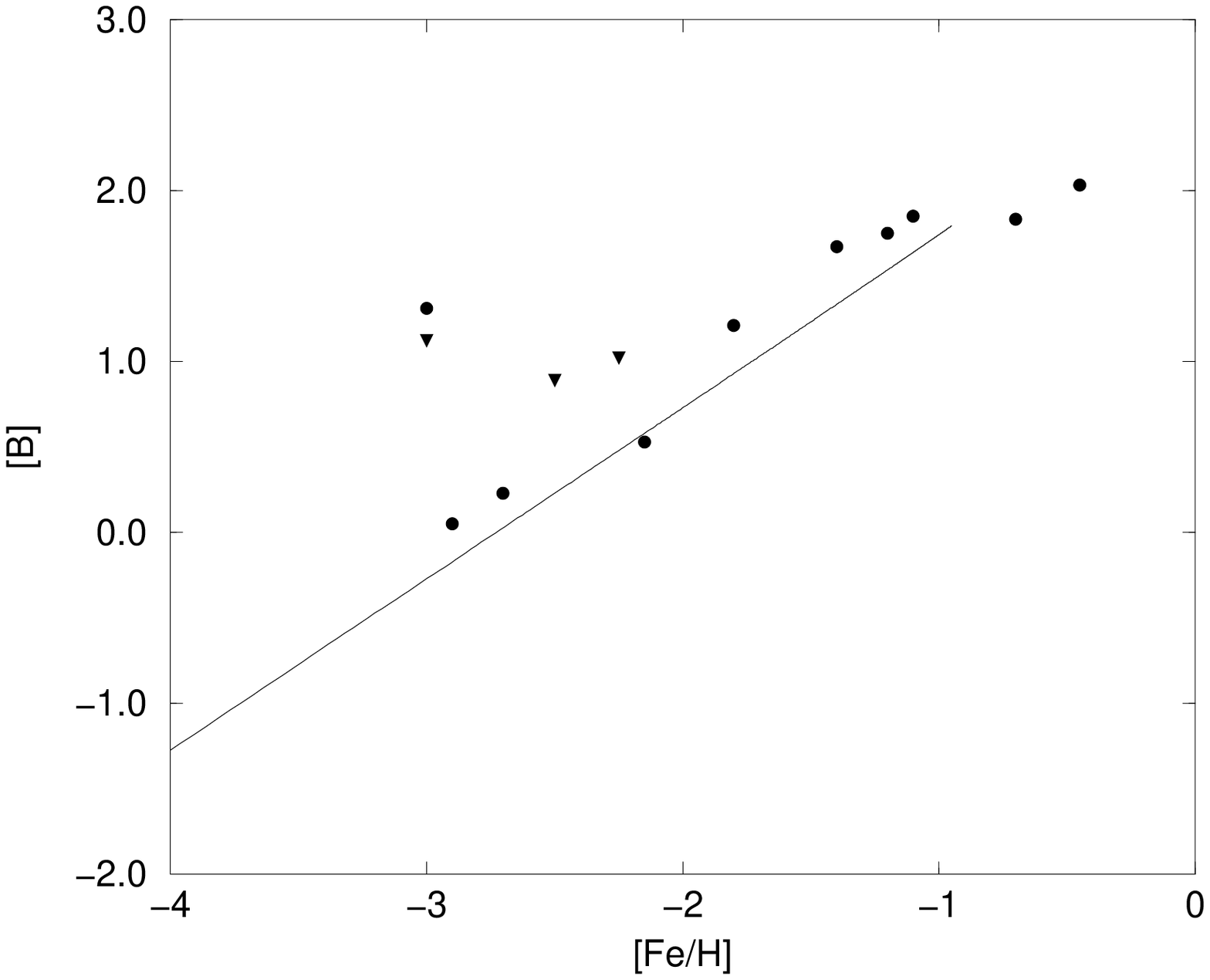}{0cm}{0}{40}{40}{-30}{-28}
\caption{\em Beryllium and boron evolution without $\nu$-process. Beryllium data: Molaro et al. (1997). Boron data: Garc\'{\i}a L\'opez et al. (1998)}
\end{figure}

 \begin{figure}[h]
\plotfiddle{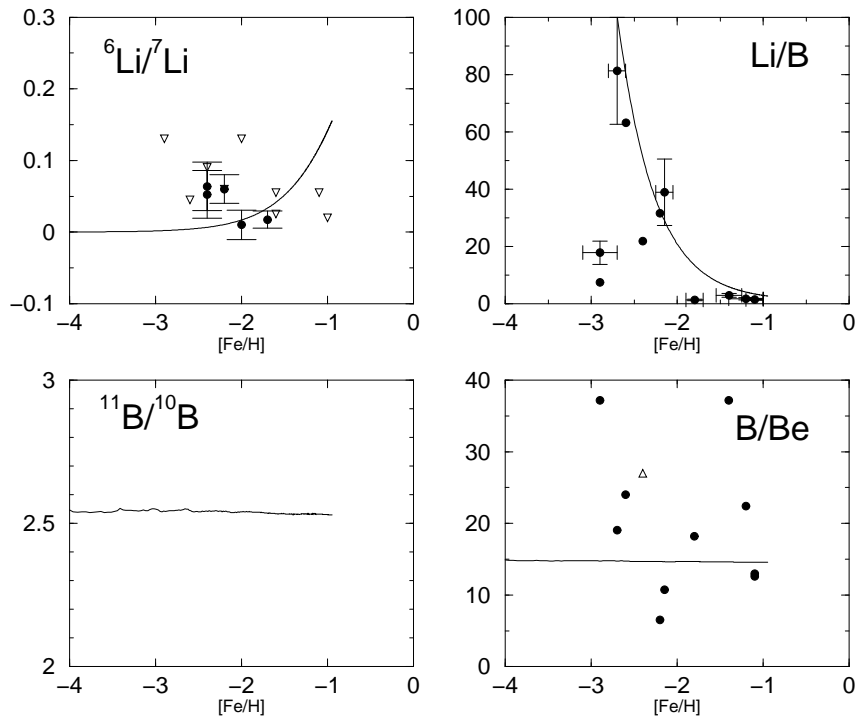}{7cm}{0}{60}{60}{-200}{-20}
\caption{\em Light-element ratios evolution without $\nu$-process. Data: Hobbs \& Thorburn (1997), Garc\'{\i}a L\'opez et al. (1998), Smith et al. (1998)}
\end{figure}
\subsection{Limits to the neutrino-induced nucleosynthesis}

In the next step, the yields of neutrino-induced nucleosynthesis theoretically calculated by Woosley \& Weaver (1995) are included. When the full yields are used ($\alpha_\nu$ = 1.0), neither the lithium plateau nor the linear boron evolution are appreciably affected (fig. 6), in spite of the production of $^7$Li and $^{11}$B by the $\nu$-process.

\begin{figure}[h]
\plotfiddle{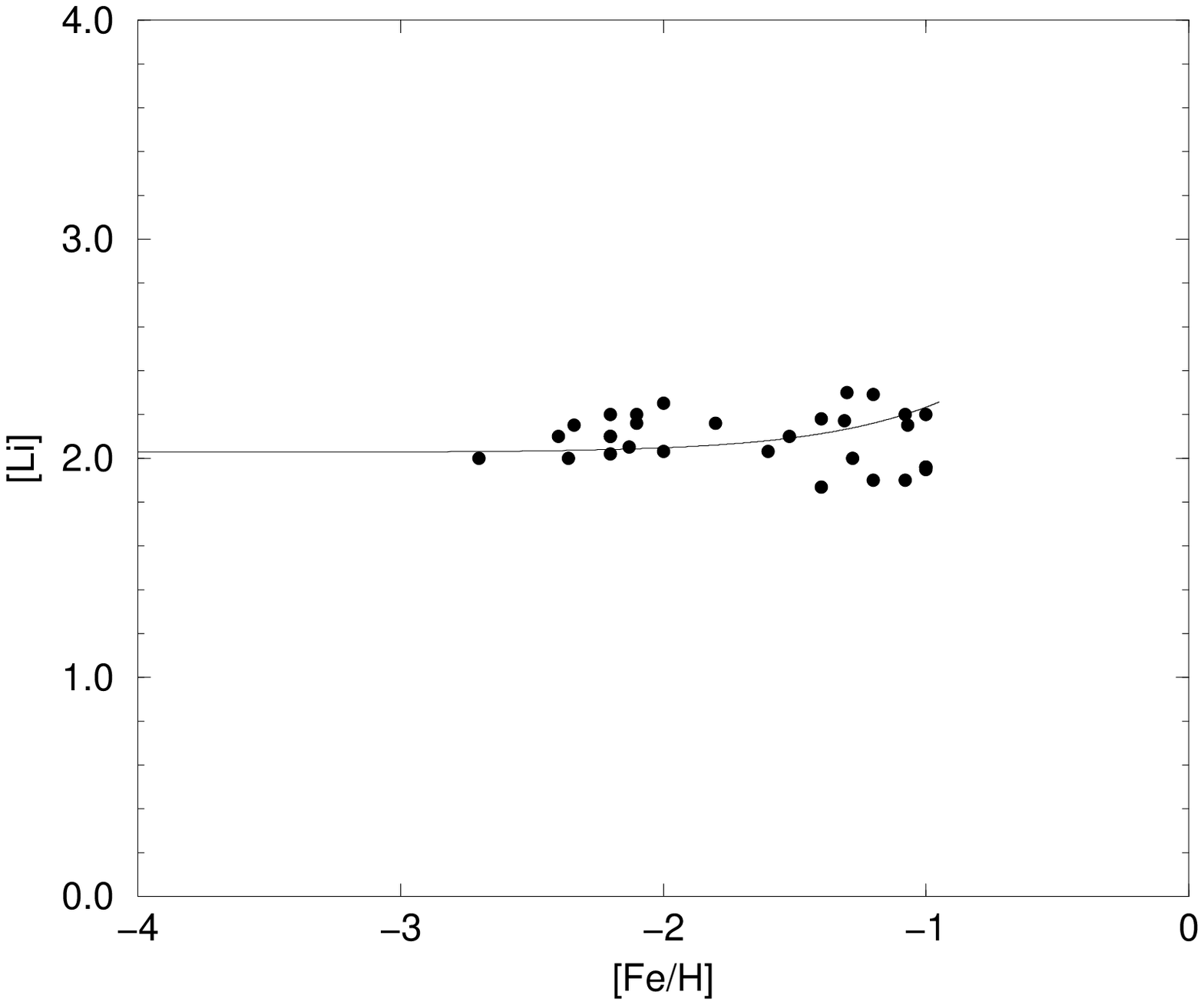}{4cm}{0}{40}{40}{-220}{-50}
\plotfiddle{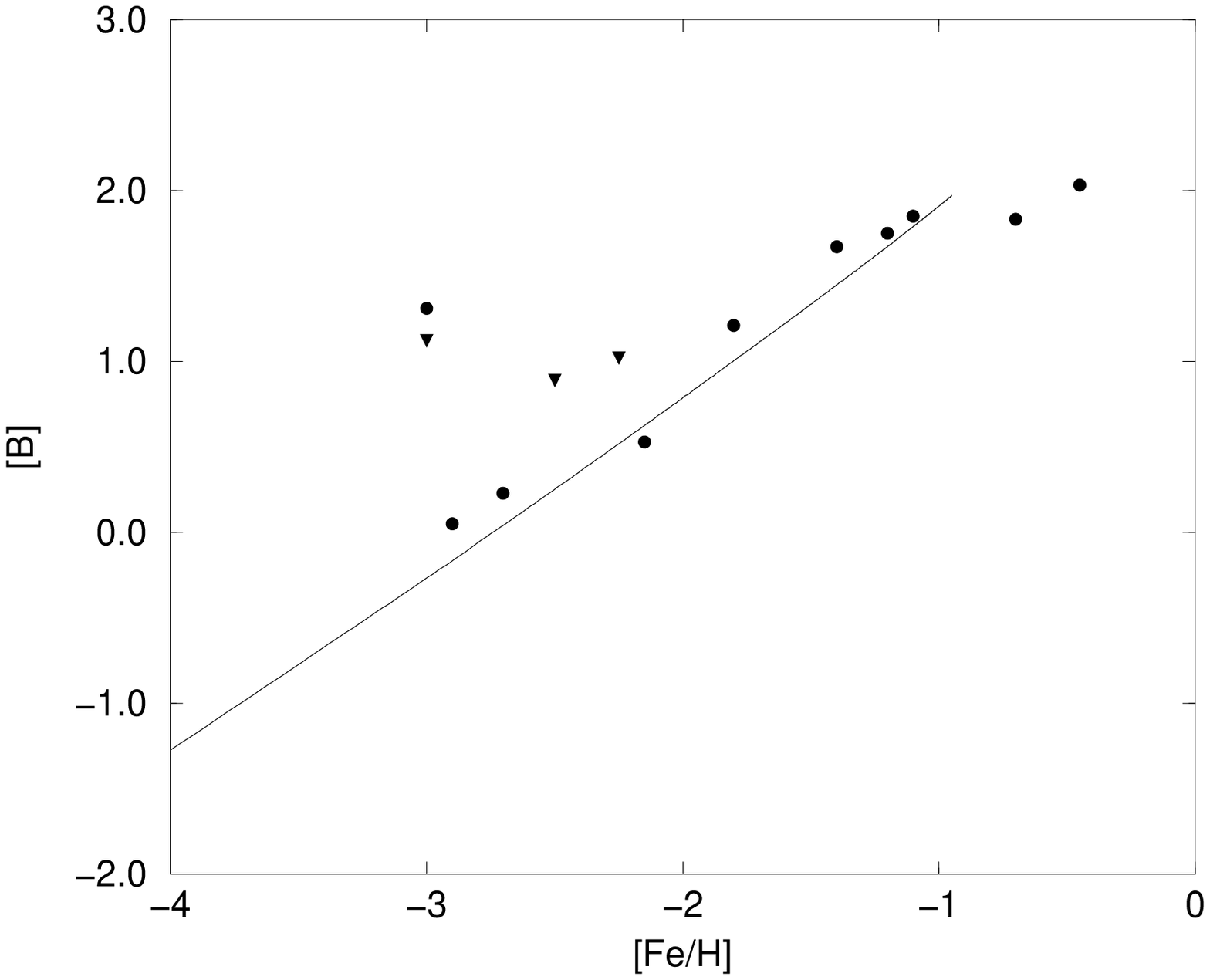}{0cm}{0}{40}{40}{-30}{-28}
\caption{\em Lithium and boron evolution including $\nu$-process ($\alpha_\nu$ = 1.0)}
\end{figure}

To reproduce the LiBeB isotopic ratios would be the main way of setting limits to the contribution of neutrino-induced nucleosynthesis to the light-element evolution in the halo. In spite of the lack of LiBeB isotopic ratios data, several facts can be considered: {\em i}) $\nu$-process doesn't affect the elements evolution; {\em ii}) if we consider that the subsequent disk evolution doesn't produce any change in the $^{11}$B/$^{10}$B ratio, a $\alpha_\nu$ = 0.85 would be necessary; {\em iii}) if this ratio changes in the disk evolution, values from $\alpha_\nu$ = 0.5 to 1.0 should be considered (fig. 7). A galactic halo+disk model (work in preparation) will restrict better this parameter.

\begin{figure}[h]
\plotfiddle{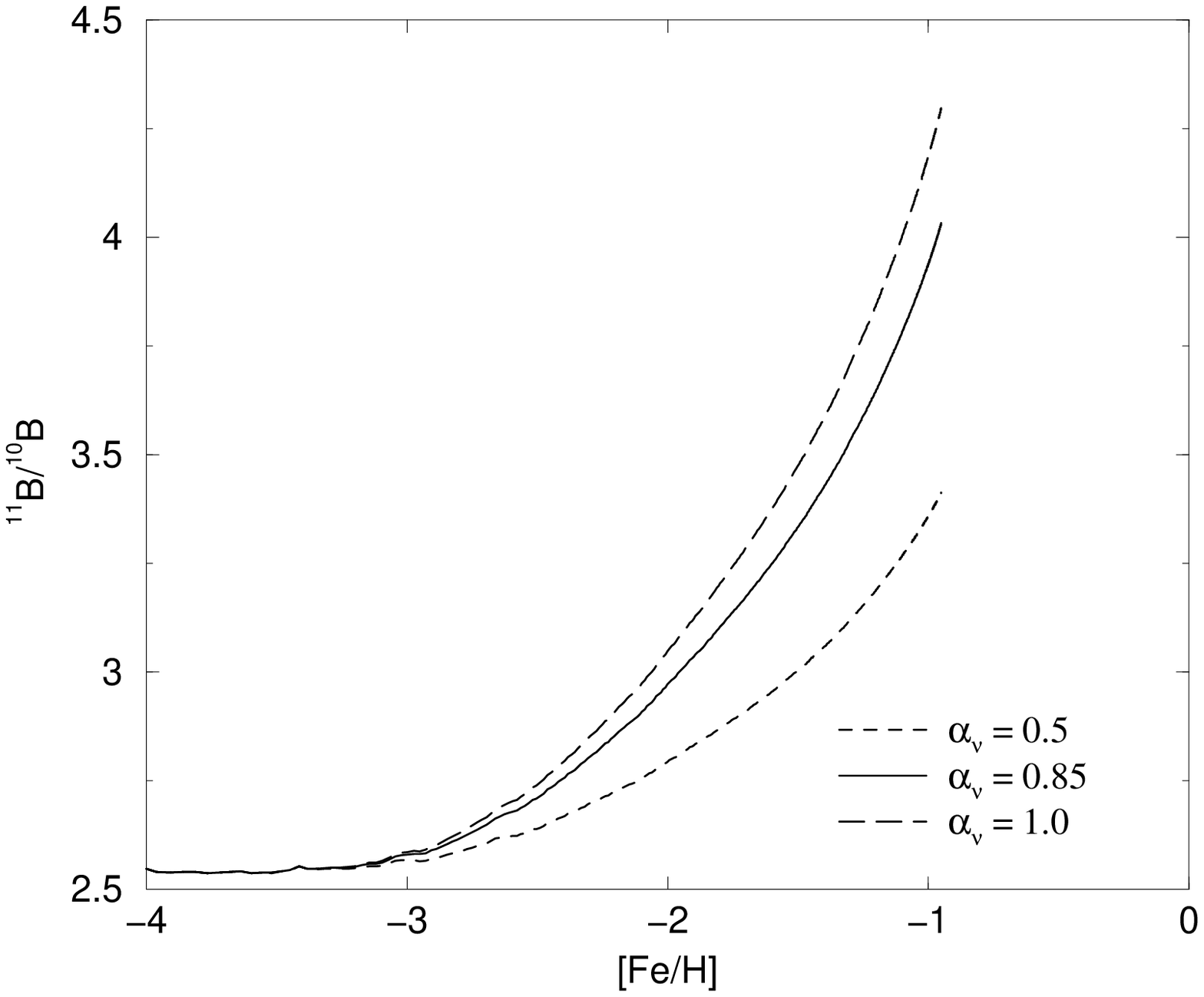}{4cm}{0}{40}{40}{-220}{-50}
\plotfiddle{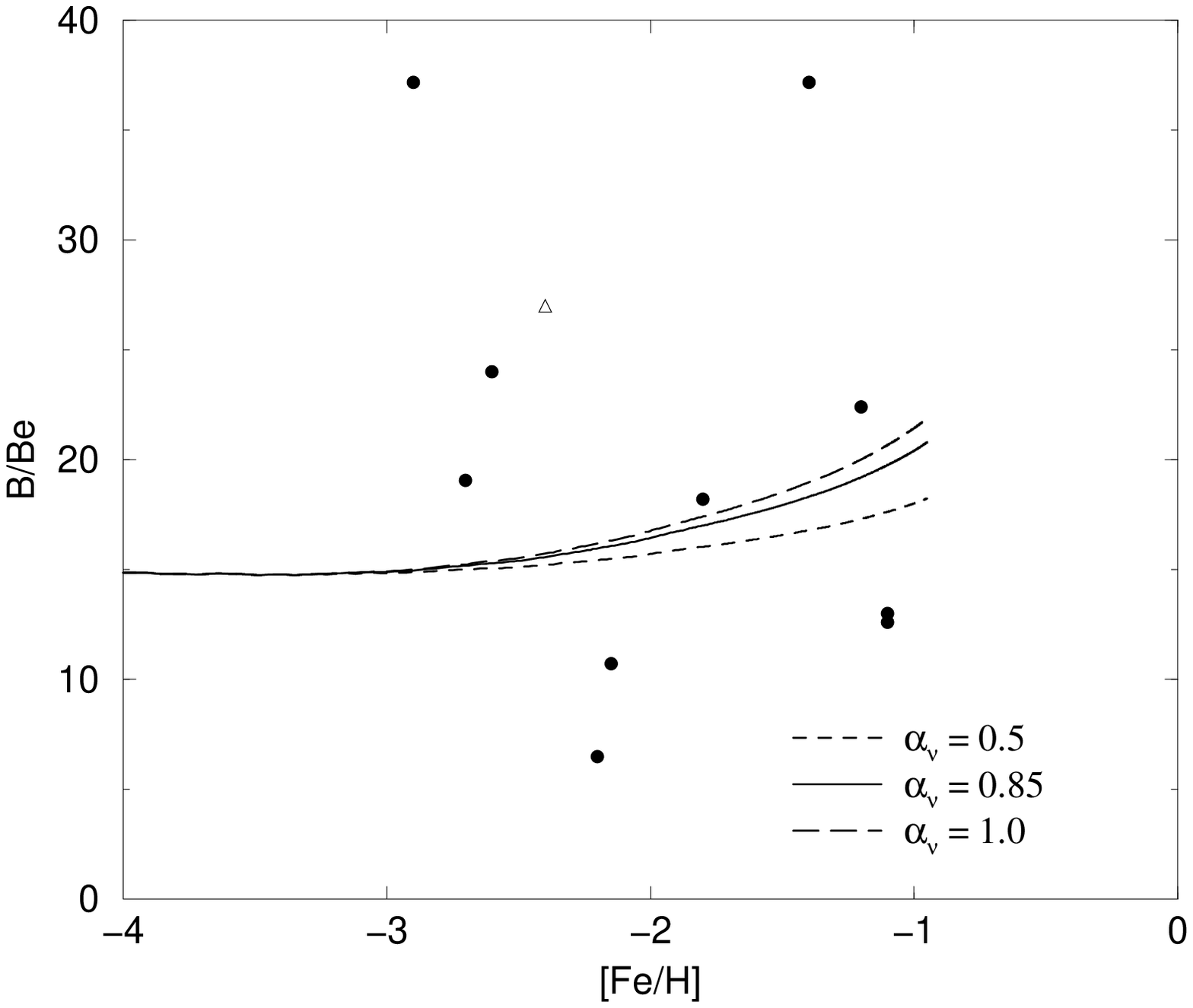}{0cm}{0}{40}{40}{-30}{-28}
\caption{\em Light-element isotopic ratios evolution including $\nu$-process}
\end{figure}

\section{Conclusions}

In the Galactic Halo, as it has been found for the light element evolution in the disk, Galactic Cosmic Ray Nucleosynthesis alone can not be the only source of LiBeB. As the existence of Low Energy Cosmic Rays is nowadays doubtful, the $\nu$-process (Woosley et al. 1990) is another possible source, mainly for $^7$Li and $^{11}$B. If neutrino-induced nucleosynthesis contributes to  LiBeB evolution, Woosley \& Weaver's yields must be revised, due to the uncertainities involved in its calculation, in order to reproduce the isotopes evolution. To constrain further the range of parameters of our model, more measurements fo the light-element ratios would be required, especially those on $^{11}$B/$^{10}$B.

\acknowledgments

We thank Dr. R. Ramaty for kindly providing us with his numerical code for cosmic-ray induced nucleosynthesis.

\end{document}